\documentclass[final,5p,times,twocolumn]{elsarticle}

\usepackage{lineno,hyperref}
\usepackage{siunitx}
\usepackage{amsmath}

\modulolinenumbers[5]

\journal{Nuclear Instruments and Methods in Physics Research A.}

\begin{document}

\begin{frontmatter}

\title{First design of a superconducting qubit for the QUB-IT experiment}


\author[bicocca,infn-bicocca]{Danilo Labranca}
\author[ing-pisa,infn-firenze]{Hervè Atsè Corti}
\author[firenze,infn-firenze]{Leonardo Banchi}
\author[ing-firenze,infn-firenze]{Alessandro Cidronali}
\author[complex]{Simone Felicetti}
\author[frascati]{Claudio Gatti}
\author[bicocca,infn-bicocca]{Andrea Giachero}
\author[bicocca,infn-bicocca]{Angelo Nucciotti}

\address[bicocca]{Department of Physics, University of Milano-Bicocca - Milano, Italy}
\address[infn-bicocca]{INFN, Sezione di Milano-Bicocca - Milano, Italy}
\address[ing-pisa]{Department of Information Engineering, University of Pisa - Pisa, Italy}
\address[infn-firenze]{INFN, Sezione di Firenze - Sesto Fiorentino (FI), Italy}
\address[firenze]{Department of Physics and Astronomy, University of Florence - Sesto Fiorentino (FI), Italy}
\address[ing-firenze]{Department of Information Engeneering, University of Florence - Firenze, Italy}
\address[complex]{Institute for Complex Systems, National Research Council (ISC-CNR) - Rome, Italy}
\address[frascati]{Laboratori Nazionali di Frascati dell’INFN - Frascati (RM), Italy}

\begin{abstract}
Quantum sensing is a rapidly growing field of research which is already improving sensitivity in fundamental physics experiments. The ability to control quantum devices to measure physical quantities received a major boost from superconducting qubits and the improved capacity in engineering and fabricating this type of devices. The goal of the QUB-IT project is to realize an itinerant single-photon counter exploiting Quantum Non Demolition (QND) measurements and entangled qubits, in order to surpass current devices in terms of efficiency and low dark-count rates. Such a detector has direct applications in Axion dark-matter experiments (such as QUAX\cite{Alesini:2020vny}), which require the photon to travel along a transmission line before being measured. In this contribution we present
the design and simulation of the first superconducting device consisting of a transmon qubit coupled to a resonator using the Qiskit-Metal software developed by IBM.
Exploiting the Energy Participation Ratio (EPR) simulation we were able to extract the circuit Hamiltonian parameters, such as resonant frequencies, anharmonicity and qubit-resonator couplings.

\end{abstract}


\end{frontmatter}


\section{Design}
\begin{figure}[t]
\begin{minipage}{.45\linewidth}
\centering  \includegraphics[height=3.5cm]{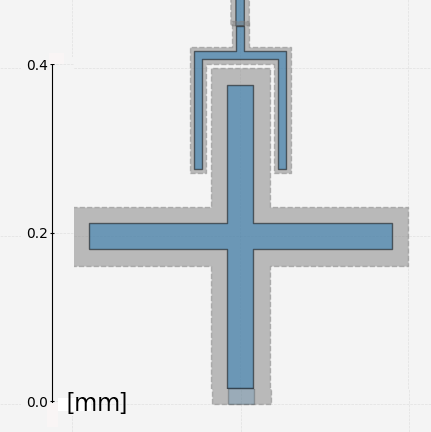}
  
  \text{(a)}
  \label{fig:design_xmon}
\end{minipage}
\centering
\begin{minipage}{.45\linewidth}
\centering
  \includegraphics[height=3.5cm]{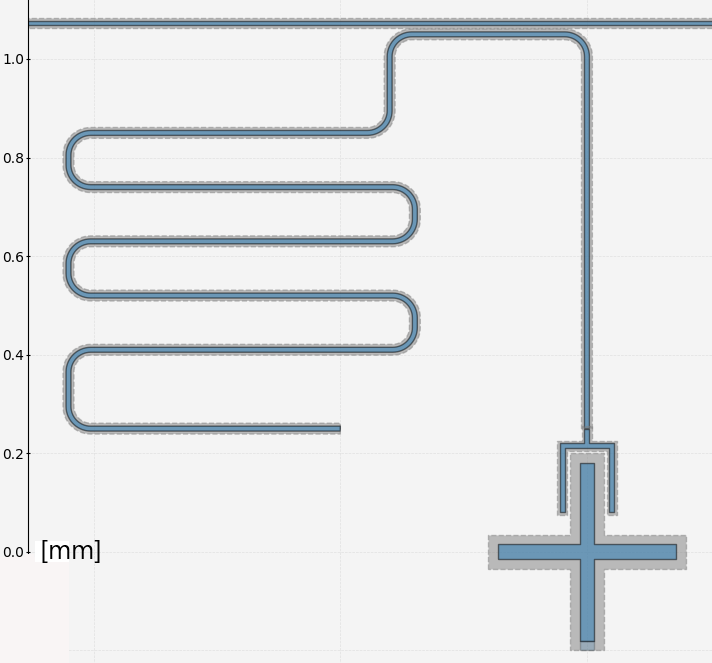}
  
  \text{(b)}
  \label{fig:design_xmonres}
\end{minipage}%

\label{fig:design}
\caption{Design rendered in Qiskit-Metal GUI. (a) Zoom on the Xmon qubit: this type of qubit consists of a Josephson Junction shunted by a cross-shaped capacitance. (b) Xmon qubit capacitively coupled to a quarter wave resonator for readout. The readout resonator is also capacitively coupled to the feedline. }
\end{figure}

\begin{figure}[t]
    \centering
    \includegraphics[height=5.1cm]{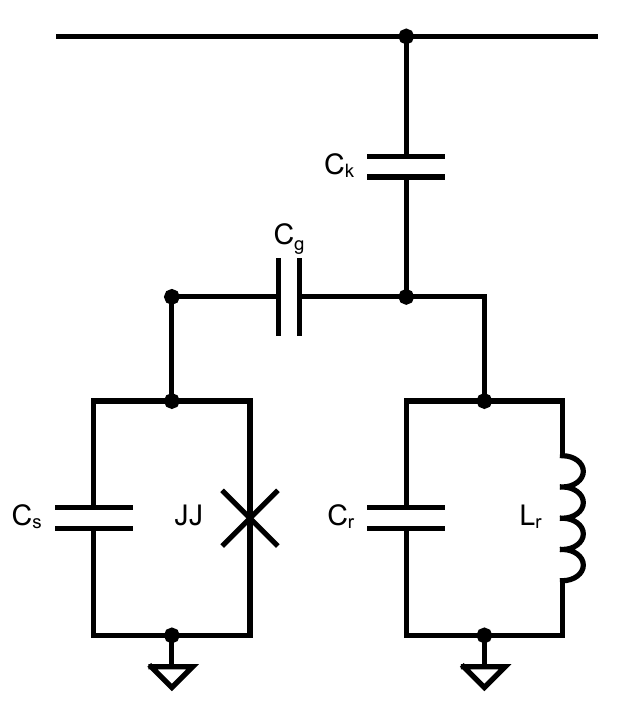}
    \caption{Lumped element circuit equivalent of our design. The resonator is represented as an LC circuit ($L_r$, $C_r$), while the qubit is a Josephson Junction shunted by a capacitance ($C_s$). $C_g$ is the qubit-resonator coupling capacitance and $C_k$ is the resonator-feedline coupling capacitance.}
    \label{fig:circuit}
\end{figure}

The design and simulation phase is fundamental in order to address the best circuit parameters for the superconducting qubit device before moving to manufacturing stage. The aim of our first design is to build a superconducting qubit that can be controlled and read through a resonator coupled to a feedline. For this chip we considered a substrate consisting of \SI{0.3}{\micro\meter} of silicon oxide on top of \SI{600}{\micro\meter} of silicon. 
\newline
The design of our circuit was developed exploiting Qiskit-Metal \cite{Qiskit_Metal} toolkit and it is shown in Fig. \ref{fig:design}. The circuit consists of an Xmon-type qubit capacitively coupled to a quarter wave readout resonator. Each electrode of the Xmon cross is \SI{30}{\micro\meter} wide and \SI{360}{\micro\meter} long. The gap between the cross and the ground plane is \SI{20}{\micro\meter} wide. The readout resonator is  \SI{5.76}{\milli\meter} long, with \SI{10}{\micro\meter} trace width and \SI{6}{\micro\meter} gap, matching a characteristic impedance $Z_0 = \SI{50}{\Omega}$. The coupling element is a \SI{130}{\micro\meter} long claw, with \SI{5}{\micro\meter} gap to the ground plane. The resonator is capacitively coupled to the feedline by a \SI{400}{\micro\meter} long coupling leg. The lumped element circuit equivalent to our design is shown in Fig. \ref{fig:circuit}.
\section{Simulations}
\begin{figure}[t]
    \centering
    \includegraphics[height=5.5cm]{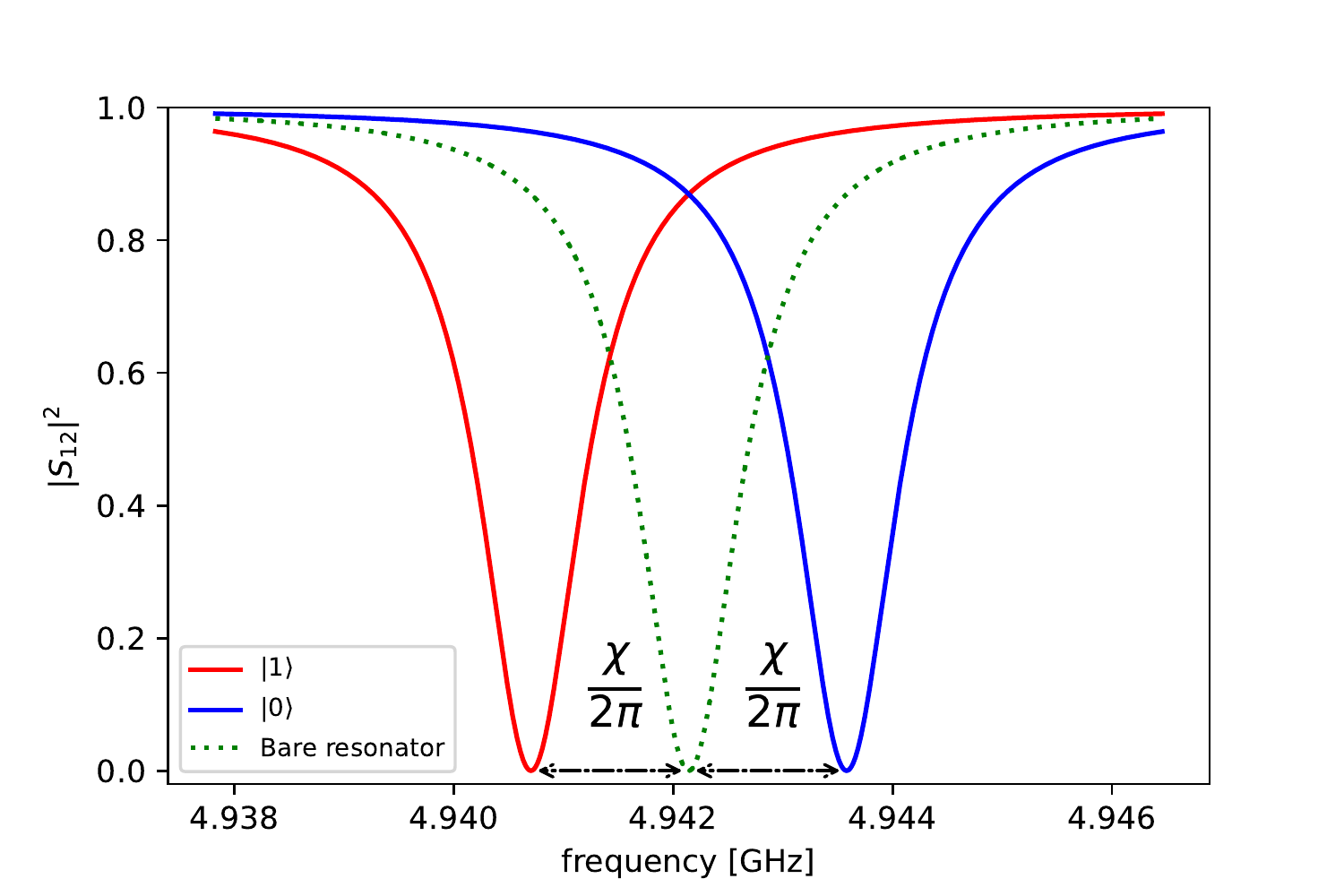}
    \caption{Feedline transmission as a function of frequency when the qubit is in the ground (blue) or excited state (red).}
    \label{fig:readout}
\end{figure}

The simulations are performed exploiting Ansys Q3D and Ansys HFSS\footnote{ANSYS Electronics Desktop 2021 R2}. The former is used to extract the capacitance values for circuit elements. The simulated capacitances are $C_s = \SI{98.19}{\femto\farad}$, for the qubit shunt capacitance, $C_g = \SI{4.40}{\femto\farad}$ and $C_k = \SI{8.62}{fF}$ for the qubit-resonator and resonator-feedline coupling capacitances respectively. The lumped element equivalent inductance and capacitance for a quarter wave resonator can be calculated as:
\begin{align}
\centering
    C_r &= \frac{\pi}{4 \omega_r Z_0} \approx \SI{499}{\femto\farad}\\
    L_r &= \frac{1}{C_r\omega_r^2} \approx \SI{2.03}{\nano\henry}
\end{align}
from \cite{pozar2011microwave}.
\newline
The chosen value for the Josephson junction inductance is $L_j = \SI{11}{\nano\henry}$, matching a critical current of $I_c = \SI{29.92}{\nano\ampere}$. With these circuit parameters we expect a Josephson energy $E_j = \SI{14.86}{\giga\hertz}$ and a capacitve energy $E_c = \SI{188.80}{\mega\hertz}$, resulting in a ratio $E_j/E_c \approx 79$ in the transmon regime. The qubit and readout resonator resonant frequencies are $\omega_{01}/2\pi = \SI{4.55}{\giga\hertz}$ and $\omega_r/2\pi = \SI{5.01}{\giga\hertz}$ respectively, leading to a detuning of $\Delta_0/2\pi \approx \SI{457}{\mega\hertz}$.
The qubit-resonator coupling strength can be calculated as (\cite{Koch_2007}):
\begin{equation}
    g_{n, n+1} = \sqrt{n + 1}\frac{2 \beta e V_{rms}}{\hbar} \left(\frac{E_j}{32 E_c}\right)^{\frac{1}{4}}
\end{equation}
with $\beta = C_g/(C_g + C_s)$, $V_{rms}=\sqrt{\hbar \omega_r / 2 C_r}$ and $e$ elementary charge, leading to $g_{01}/ 2 \pi = \SI{47.38}{\mega\hertz}$. We can then calculate the total dispersive shift as (\cite{Koch_2007}):
\begin{equation}
    \chi = \chi_{01} - \frac{\chi_{12}}{2}
\end{equation}
where $\chi_{ij} = g_{ij}^2/(\omega_{ij} - \omega_r)$.
The total dispersive shift value is $\chi/2\pi = \SI{-1.44}{\mega\hertz}$, being larger than the resonator resonance width $\kappa/2\pi = \SI{1.12}{\mega\hertz}$ allows the qubit state readout. Fig. \ref{fig:readout} shows the expected transmission on the feedline as a function of frequency for the qubit in the ground or excited state.
\newline
An estimate of the qubit relaxation time can be calculated as follows (\cite{Blais_2004}):
\begin{equation}
    T_1 = \frac{\Delta_0^2}{g_{01}^2}\frac{Q}{\omega_r} \approx \SI{13}{\micro\second}
\end{equation}
where $Q \approx 4432$ (calculated as in \cite{G_ppl_2008}) is the resonator quality factor.
\newline
Ansys HFSS is used to perform the eigenmode simulation in order to compute the resonant frequencies of the circuit and exploit the Energy Participation Ratio (EPR) simulation \cite{EPR,pyEPR}. From the EPR analysis we can extract the qubit and resonator frequencies, anharmonicities and the total dispersive shift. The simulated qubit and resonator frequencies are $\omega_{01}^{EPR}/2 \pi = \SI{4.43}{\giga\hertz}$ and $\omega_{r}^{EPR}/2 \pi = \SI{5.17}{\giga\hertz}$ respectively. These values match the expected frequencies within $2.6\%$ and $3.2\%$ respectively. The qubit anharmonicity extracted from the EPR simulation is $\alpha^{EPR} = \SI{-193.43}{\mega\hertz}$ and matches the expected $E_c$ within $2.5\%$. The simulated value for the total dispersive shift is $\chi^{EPR}/2\pi = \SI{-1.37}{\mega\hertz}$, within $4.9\%$ the expected value.
\section{Conclusions}
We were able to simulate one possible design for the first chip fabrication, comparing the results with the expected circuit features. The simulations agree with the calculated values within a few percents. It should be noted that longer and more accurate simulations could further reduce these gaps.
\newline
Before moving to the manufacturing stage, a similar design with inductive resonator-feedline coupling will be evaluated, as well as different qubit and Josephson junction parameters. Higher values for the anharmonicity and longer qubit relaxation time will be pursued by accurately tuning couplings and frequencies. Further parameter tuning is ongoing to optimize the qubit readout.
A simulation-data comparison will be performed as soon as the first chip is produced in order to validate the design procedure.


\bibliography{proceeding.bib}

\begin{thebibliography}{1}
\expandafter\ifx\csname url\endcsname\relax
  \def\url#1{\texttt{#1}}\fi
\expandafter\ifx\csname urlprefix\endcsname\relax\def\urlprefix{URL }\fi
\expandafter\ifx\csname href\endcsname\relax
  \def\href#1#2{#2} \def\path#1{#1}\fi

\bibitem{Alesini:2020vny}
D.~Alesini, et~al., {Search for invisible axion dark matter of mass
  m$_a=43~\mu$eV with the QUAX--$a\gamma$ experiment}, Phys. Rev. D 103~(10)
  (2021) 102004.
\newblock \href {http://arxiv.org/abs/2012.09498} {\path{arXiv:2012.09498}},
  \href {http://dx.doi.org/10.1103/PhysRevD.103.102004}
  {\path{doi:10.1103/PhysRevD.103.102004}}.

\bibitem{Qiskit_Metal}
Z.~K. Minev, et~al., {Qiskit Metal: An Open-Source Framework for Quantum Device
  Design {\&} Analysis} (2021).
\newblock \href {http://dx.doi.org/10.5281/zenodo.4618153}
  {\path{doi:10.5281/zenodo.4618153}}.

\bibitem{pozar2011microwave}
D.~Pozar, Microwave Engineering, 4th Edition, Wiley, 2011.

\bibitem{Koch_2007}
J.~Koch, et~al., Charge-insensitive qubit design derived from the cooper pair
  box, Physical Review A 76~(4).
\newblock \href {http://dx.doi.org/10.1103/physreva.76.042319}
  {\path{doi:10.1103/physreva.76.042319}}.

\bibitem{Blais_2004}
A.~Blais, et~al., Cavity quantum electrodynamics for superconducting electrical
  circuits: An architecture for quantum computation, Phys. Rev. A 69 (2004)
  062320.
\newblock \href {http://dx.doi.org/10.1103/PhysRevA.69.062320}
  {\path{doi:10.1103/PhysRevA.69.062320}}.

\bibitem{G_ppl_2008}
M.~Göppl, et~al., Coplanar waveguide resonators for circuit quantum
  electrodynamics, Journal of Applied Physics 104~(11) (2008) 113904.
\newblock \href {http://dx.doi.org/10.1063/1.3010859}
  {\path{doi:10.1063/1.3010859}}.

\bibitem{EPR}
Z.~K. Minev, et~al., Energy-participation quantization of josephson circuits
  (2020).
\newblock \href {http://dx.doi.org/10.48550/arxiv.2010.00620}
  {\path{doi:10.48550/arxiv.2010.00620}}.

\bibitem{pyEPR}
Z.~K. Minev, et~al., {pyEPR: The energy-participation-ratio (EPR) open-source
  framework for quantum device design} (May 2021).
\newblock \href {http://dx.doi.org/10.5281/zenodo.4744447}
  {\path{doi:10.5281/zenodo.4744447}}.

\end{thebibliography}

\end{document}